\documentclass[aps,prd,10pt,twocolumn,floatfix,superscriptaddress,preprintnumbers,nofootinbib,notoccite,notitlepage]{revtex4-1}

\usepackage{graphicx}
\usepackage{dcolumn}
\usepackage{footmisc}
\usepackage{bm}
\usepackage[dvipsnames,table]{xcolor}
\usepackage{hyperref}
\usepackage{float}
\usepackage{amsmath, amssymb}
\usepackage{tikz-feynman}
\usepackage{mdframed}
\usepackage{multirow}
\usepackage{tcolorbox}
\usepackage{orcidlink}

\usepackage[ulem=normalem, commandnameprefix=ifneeded]{changes}

\begin{document}
\rightline{}
\raggedbottom

\preprint{}

\title{Direct Detection of Millicharged Particles from Supernovae}
\author{Yanou Cui, \orcidlink{0000-0003-4322-9246}} 
\email{yanou.cui@ucr.edu}
\affiliation{Department of Physics and Astronomy, University of California, Riverside, CA 92521, USA}
\author{Fengyi Li, \orcidlink{0009-0005-6595-2197}}
\email{fli044@ucr.edu}
\affiliation{Department of Physics and Astronomy, University of California, Riverside, CA 92521, USA}
\author{Xiaolin Qi, \orcidlink{0009-0002-9747-0334}}
\email{xiaolinq76@vt.edu}
\affiliation{Center for Neutrino Physics, Department of Physics,
Virginia Tech, Blacksburg, Virginia 24061, USA}
\author{Ian M.~Shoemaker,\orcidlink{0000-0001-5434-3744}}
\email{shoemaker@vt.edu}
\affiliation{Center for Neutrino Physics, Department of Physics,
Virginia Tech, Blacksburg, Virginia 24061, USA}
\author{Yu-Dai Tsai,\orcidlink{0000-0002-5763-5758}}
\email{y.tsai@sheffield.ac.uk}
\email{yudaitsai.academic@gmail.com}
\affiliation{The University of Sheffield, Sheffield S3 7RH, UK}
\affiliation{The University of Manchester, Manchester M13 9PL, UK}
\affiliation{Los Alamos National Laboratory (LANL), Los Alamos, NM 87545, USA}

\begin{abstract}
This work proposes a new terrestrial probe for millicharged particles (mCPs) and demonstrates promising discovery prospects. mCPs can be copiously produced in core-collapse supernovae (SNe), and a fraction may escape, travel to Earth and yield distinct signals. The mCP mass induces a time-of-flight (ToF) delay relative to the SN neutrino burst,
opening a clean search window after the neutrino signal has passed. We compute the mCP-induced electron-recoil signals at XENONnT, JUNO, DUNE, and Hyper-Kamiokande for benchmark SN scenarios, 
and find that for $\varepsilon = 10^{-9}$ and sub-MeV to MeV-scale masses, more than 10 events per year can be detected. This search can improve upon existing SN cooling bound on $\varepsilon$ by up to an order of magnitude.
\end{abstract}

\maketitle

\section{Introduction}\label{sec:intro}

Electric charge appears quantized in every established particle of nature, yet this quantization is not dictated by the low-energy gauge structure of the Standard Model (SM). The possibility of particles carrying a tiny, non-standard electric charge is theoretically motivated and inspired experimental quests testing charge quantization~\cite{Dirac:1931kp,Pati:1973uk}. Millicharged particles (mCPs) realize this possibility in one of its simplest forms. Such states arise naturally when ordinary electromagnetism communicates with a hidden Abelian gauge sector, for example through kinetic mixing~\cite{Holdom:1985ag,Holdom:1986eq} or Stueckelberg mass mixing~\cite{Cheung:2007ut,Feldman:2007wj}. They are also motivated by broader hidden-sector and string-theoretic constructions, where extra $\mathrm{U}(1)$ gauge symmetries are common and small effective electromagnetic charges can emerge for otherwise dark particles~\cite{Wen:1985qj}. Therefore, searches for mCPs probe not only a new species of light particle, but also the fundamental theory, the organizing principles, and the mathematical structure of nature~\cite{Essig:2013lka,Battaglieri:2017aum,Gori:2022vri,Lee:2026djo}.

A fermionic mCP $\chi$ interacts with SM photon through
\begin{equation}
\mathcal{L}_{\rm int}=\varepsilon e\, A_{\mu}\, \bar{\chi}\gamma^\mu\chi,
\label{eq:Lint}
\end{equation}
where $e$ is the SM electromagnetic coupling, $\varepsilon < 1$ is the millicharge parameter of $\chi$, $A_{\mu}$ is the SM photon, and $\gamma^{\mu}$ is a Dirac gamma matrix. This effective interaction can arise from assigning a small hypercharge to $\chi$, thereby inducing couplings to both the photon and the $Z$ boson after the electroweak symmetry breaking~\cite{zcoupling}.

Various searches for mCPs have been conducted, resulting in constraints on the mCP parameter space 
$(\varepsilon,\, m_{\chi})$. Such searches include: laboratory experiments~\cite{Ahlers:2007qf,DellaValle:2014xoa,Magill:2018tbb,Berlin:2018bsc,Kelly:2018brz,Foroughi-Abari:2020qar,Oscura:2023qch,SENSEI:2023gie,Tsai:2024wdh,Essig:2024dpa,Berlin:2024dwg,Berlin:2025hjs,Berlin:2025btf}, as well as observations in astrophysics~\cite{Davidson:1993sj,Chang:2018rso,Harnik:2020ugb,Plestid:2020kdm,ArguellesDelgado:2021lek,Berlin:2021kcm,Li:2024pcp,Wu:2024iqm,Fiorillo:2024upk}, and cosmology~\cite{Davidson:2000hf,Gan:2023jbs,Iles:2024zka}.
In this work, we focus in particular on mCPs sourced by core-collapse supernovae (SNe), which are among the most extreme environments in the universe. The collapsing core reaches temperatures of tens of MeV with densities exceeding $10^{14}\; \mathrm{g/cm^3}$, enabling efficient production of not only SM particles but also beyond-the-SM (BSM) particles with very weak couplings to ordinary matter. Thus SNe can serve as powerful laboratories for probing such BSM states, including mCPs.

Two types of SN-derived constraints are particularly relevant to this work and are taken into account here: the SN cooling bound and the low-energy supernova (LESN) constraint. The former requires any new energy-loss channel to be less efficient than standard neutrino emission, since otherwise it would shorten the neutrino burst observed from SN1987A. The latter requires that any additional energy deposited from the core into the supernova mantle does not exceed the observed explosion energy~\cite{Li:2024pcp,Caputo:2022mah,Chang:2018rso}.

This work focuses on the regime where mCPs can escape the supernova core freely and travel to Earth, and puts forth a new type of signal for mCP discovery, using neutrino or dark matter detectors. Notably, the mCP mass induces a time-of-flight (ToF) delay relative to the neutrino burst from the same SN explosion. That neutrino burst triggers the search across the global detector network, after which the mCP signal distinguishes itself through its extended time profile and its distinct electron recoil spectrum.
Interestingly, the projected sensitivity of our proposed search reaches parameter regions in which mCPs can arise as dark matter candidates through freeze-in production. Nevertheless, for mCPs in this parameter region to account for the full amount of dark matter while being compatible with existing constraints, a low reheating temperature or additional model-building may be required~\cite{Dvorkin:2019zdi,Gan:2023jbs,Iles:2024zka,Boddy:2024vgt}.

The remainder of this paper is organized as follows. In Sec.~\ref{sec:production} we review mechanisms of mCP production in the supernova core. In Sec.~\ref{sec:propagation} we discuss their propagation to Earth, with particular emphasis on the ToF delay and magnetic-field effects. In Sec.~\ref{sec:detection}, we present the detection strategy and specifics of benchmark detectors. The main results are given in Sec.~\ref{sec:results}, with conclusions in Sec.~\ref{sec:conclusions}.\\

\section{Millicharged Particle Production in Supernovae}\label{sec:production} 

\vspace{6pt}
\textbf{\textit{Supernova Model.}} A core-collapse supernova begins when a massive star exhausts its nuclear fuel. The iron core, no longer supported by fusion, collapses under gravity within milliseconds, reaching nuclear densities and forms a hot, compact proto-neutron star (PNS). A shock wave launches outward through the stellar envelope, while the PNS core settles into approximate thermal equilibrium. 

The conditions inside the PNS core define the production environment for mCPs. This work adopts the one-zone model of Ref.~\cite{Li:2024pcp}, based on a snapshot of the PNS at approximately one second post-bounce. The core has radius $R_c = 12.9$~km, temperature $T_c = 30$~MeV, nuclear density $\rho_c = 3 \times 10^{14}~\mathrm{g/cm^3}$, proton fraction $Y_p = 0.15$, and electron chemical potential $\mu_e \simeq 167$~MeV. At these temperatures and densities, the core is capable of producing mCPs through interactions enabled by the coupling in Eq.~\eqref{eq:Lint}.
These PNS core parameters, used in the SN cooling literature~\cite{Chang:2018rso,Raffelt:1996wa} and the LESN literature~\cite{Li:2024pcp,Caputo:2022mah}, are adopted here as well.

The reference distance between the SN and Earth for our analysis is chosen to be $D = 1$~kpc, representative of the potential next Galactic supernova. The Galactic core-collapse supernova rate is $\sim1-3$ per century~\cite{SNEWS:2021ezc}. Betelgeuse, a nearby red supergiant and supernova progenitor candidate at $D \simeq 0.17$~kpc~\cite{Joyce:2020vbx}, serves as a more optimistic benchmark. As the event number approximately scales as $1/D^2$, a more distant supernova such as SN1987A at $D \simeq 50$~kpc yields a strongly suppressed signal.

In an SN, mCPs are produced primarily through three channels: plasmon decay, $e^{+}e^{-}$ annihilation, and proton bremsstrahlung. We review these channels below.

\vspace{6pt}
\textbf{\textit{Plasmon Decay.}} Plasmons, photons that acquire an effective mass in a dense plasma, can decay into feebly interacting particles such as the mCPs, i.e., $\gamma^* \to \chi\bar{\chi}$. This process dominates mCP production at low masses and rapidly vanishes at higher masses due to the kinematic threshold, when plasmons in the SN core lack sufficient energy to produce heavy mCP pairs. The corresponding production rate is given in Appendix~\ref{app:prod_rate}.

\vspace{6pt}
\textbf{\textit{$e^+e^-$ Annihilation.}} The SN core temperature $T_c = 30$~MeV sustains a substantial 
positron population through thermal processes~\cite{Li:2024pcp}, opening a second production channel:
$e^+e^- \to \gamma^* \to \chi\bar{\chi}$ through a virtual photon.
For $m_\chi$ below the characteristic thermal and in-medium scales of the core, the annihilation spectrum depends only weakly on $m_\chi$, because the available center-of-mass energies are mainly set by $T_c$, $\mu_e$, and the in-medium electron mass. The explicit $m_\chi$ dependence enters through the final-state phase space and the production threshold, and becomes important only once $m_\chi$ approaches the typical energies of the annihilating $e^\pm$ pair. This channel dominates above $m_\chi \sim 50$~MeV~\cite{Li:2024pcp}. The cross section and production rate of this channel are shown in Appendix~\ref{app:prod_rate}.

\vspace{6pt}
\textbf{\textit{Proton Bremsstrahlung.}} Proton bremsstrahlung provides a third mCP production channel in the supernova core, arising from virtual photon emission during nucleon-nucleon scattering,
$
np \to np\gamma^\ast \to np\chi\bar{\chi}.
$ 
Previous analyses~\cite{Davidson:2000hf,Chang:2018rso,Li:2024pcp,Bailloeul:2025fde} find proton bremsstrahlung to be an important mCP production channel over the mass range from a few MeV to $\sim 50$ MeV. However, proton bremsstrahlung has notable uncertainties and remains an active area of study: its rate depends sensitively on nuclear-scattering inputs and on the treatment of the dense nuclear medium, making a consistent implementation more model-dependent than the plasmon-decay and $e^+e^-$ annihilation channels. These two other channels already cover the relevant mass range, with plasmon decay dominating at low mass and $e^+e^-$ annihilation at high mass. Therefore, we choose to omit proton bremsstrahlung in our baseline analysis. The resulting projections are thus conservative: including bremsstrahlung would increase the mCP flux and extend the reach to even smaller $\varepsilon$ in the intermediate-mass band. The core results of this work are robust against the inclusion or details of this channel.

\section{Propagation to Earth}\label{sec:propagation}

mCPs propagate through the SN mantle and the interstellar medium before reaching detectors on Earth. This section discusses three aspects of this propagation: the escape from the SN core through the mantle, the ToF delay due to the mCP mass, and the potential deflection by the Galactic magnetic field.

\vspace{6pt}
\textbf{\textit{Free Streaming and Mantle Propagation.}} 

The calculation of mCP production is detailed in Appendix~\ref{app:prod_rate}, where it is assumed that mCPs escape the SN core without rescattering. This condition can be satisfied if the mCP mean free path $\lambda_{\rm mfp}$ exceeds the core radius $R_c$. In the relativistic limit, $\chi$-$e$ Coulomb scattering cross section can be estimated as $\sigma_{\chi e} \sim \varepsilon^2 \alpha^2 / T_c^2$, and thus the condition of $\lambda_{\rm mfp} > R_c$ is met for $\varepsilon \lesssim 10^{-8}$ and $m_\chi \lesssim 100$~MeV~\cite{Chang:2018rso}, which includes the parameter space of interest in this work.

After leaving the core, mCPs traverse the SN mantle. Coulomb scattering off mantle protons transfers a fraction of the mCP kinetic energy to the shock-heated matter, which is the effect used to set the LESN bound~\cite{Li:2024pcp,Caputo:2022mah}. With $\varepsilon \sim 10^{-9}$, the per-mCP energy loss in the mantle is estimated to be less than $1\%$ of the original mCP energy across the relevant mass range. Therefore, its effect is safely negligible.

\vspace{6pt}
\textbf{\textit{Magnetic Field Effects.}} 
 Since mCPs carry a small electric charge, Galactic magnetic fields can potentially deflect them. To assess the importance of this effect, we estimate the corresponding Larmor radius,
\begin{align}
    r_L &= \frac{\gamma m_\chi v}{\varepsilon e B}
    = \frac{E_\chi}{\varepsilon e B}
    \left[1-\left(\frac{m_\chi}{E_\chi}\right)^2\right]^{1/2} \nonumber\\
    &\simeq 10~{\rm pc}
    \left(\frac{E_\chi}{10~{\rm MeV}}\right)
    \left(\frac{10^{-9}}{\varepsilon}\right)
    \left(\frac{10^{-6}~{\rm G}}{B}\right),
\end{align}
where $\gamma m_\chi v$ is the mCP relativistic momentum and $B$ is the Galactic magnetic-field strength.

A full treatment of mCP propagation in Galactic magnetic fields is nontrivial and beyond the scope of this work. A simplified assumption that mCPs propagate along straight trajectories can be justified in realistic scenarios, and is adopted in our analysis. As a proof of principle, one may consider a natural UV completion of the coupling in Eq.~\eqref{eq:Lint} that invokes a dark \(U(1)\) with a light gauge boson \(A'\), where mCPs acquire an effective SM electric charge through kinetic mixing with the SM photon. For \(A'\) mass in the range \(10^{-24}~{\rm eV} \lesssim m_{A'} \lesssim 10^{-11}~{\rm eV}\) the interaction is long-ranged on neutron-star scales but Yukawa-suppressed on scales beyond the Larmor radius estimated above. In this case, mCPs respond as ordinary charged particles inside the supernova core while magnetic deflection during Galactic propagation is negligible (assuming $B=\mu G$, $E_{\chi} = 10$ MeV, and $\varepsilon = 10^{-9}$). 

\vspace{6pt}
\textbf{\textit{ToF Delay.}} 
A massive particle travels slower than light. An mCP with mass 
$m_\chi$ and energy $\bar{E}_\chi$ arrives at Earth with a ToF delay $\Delta t$ 
relative to a massless particle given by:

\begin{align}
\Delta t &= \frac{D}{v} - \frac{D}{c} 
= \frac{D}{c}\left(\frac{1}{\sqrt{1-(m_\chi/\bar{E}_\chi)^2}}-1\right) \nonumber \\
&\simeq 60~\text{days} 
\left(\frac{m_\chi}{1~\text{MeV}}\right)^2
\left(\frac{100~\text{MeV}}{\bar{E}_\chi}\right)^2
\left(\frac{D}{1~\text{kpc}}\right)
\label{eq:time_delay}
\end{align}
where $v$ is the speed of the mCP.

Based on Eq.~\ref{eq:time_delay}, we can see that in a given time window, heavier mCPs require a higher energy to travel the same distance compared to lighter ones. For a fixed $m_{\chi}$, the mCP energy uniquely determines the ToF. Given an experimental observation time window, this relation determines the range of mCP energies that can reach the detector within that period. As an example, for a supernova at 1 kpc and $m_{\chi} = 1~\rm MeV$, only mCPs with $\bar{E}_{\chi} \gtrsim 40.5~\mathrm{MeV}$ arrive within one year after the neutrino burst.

Consequently, mCP signals spread over time, unlike the neutrino burst which lasts only $\sim$10~s. This feature has both advantages and disadvantages for detection. The time-dependent signature can help distinguish mCP signals from background neutrinos. On the other hand, the temporal spread makes it challenging for detection, since the differential event number at any given time is smaller (see App.~\ref{app:event_counts}).

\section{Detection}
\label{sec:detection}

\vspace{6pt}
\textbf{\textit{The signal of mCP-electron scattering.}} After arriving at Earth, the mCPs scatter off electrons in terrestrial 
detectors via elastic scattering. The differential cross section is ~\cite{Essig:2024dpa}:
\begin{equation}
\frac{d\sigma}{dE_e} = \pi \alpha^2 \varepsilon^2\,
\frac{2E_\chi^2 m_e + E_e^2 m_e 
- E_e\left(m_\chi^2 + m_e(2E_\chi + m_e)\right)}
{E_e^2\left(E_\chi^2 - m_\chi^2\right)m_e^2},
\end{equation}
where $E_e$ is the electron recoil energy, $m_e$ is the electron mass. The differential event number in a detector with $N_T$ target 
electrons is then: 
\begin{equation}
\frac{dN}{dE_e} = \Delta t_{\rm SN} \times N_T \int_{E_{\rm min}^{\mathrm{eff}}}^{\infty} dE_\chi\,
\frac{d\phi_\chi}{dE_\chi}\,\frac{d\sigma}{dE_e}, \label{diff_rate}
\end{equation}
where $E_{\rm min}^{\mathrm{eff}}$ is the effective minimum mCP energy defined as:
\begin{equation}
E_{\mathrm{min}}^{\mathrm{eff}} = \max\left(E_{\mathrm{min}},\, \bar{E}_{\chi}(\Delta t_{\rm obs})\right). \label{Emin_eff}
\end{equation}

Here, $\bar{E}_{\chi}(\Delta t)$ denotes the mCP energy associated with a ToF delay $\Delta t$, obtained by inverting the relation $\Delta t(\bar{E}_\chi)$ in Eq.~\eqref{eq:time_delay} to express $\bar{E}_{\chi}$ as a function of $\Delta t$. For cumulative event counts, $\Delta t_{\rm obs}$ denotes the observation window. $E_{\mathrm{min}}$ is the minimal mCP energy required to produce a given recoil energy $E_{e}$ in the detector and can be expressed as~\cite{Cappiello:2019qsw}:
 
\begin{equation}
E_{\mathrm{min}} = \left(\frac{E_{e}}{2}- m_{\chi}\right)\left(1 \pm 
\sqrt{1+ \frac{2 E_{e}}{m_{e}}\frac{(m_{e}+ m_{\chi})^{2}}{(2m_{\chi}-E_{e})^{2}}}\right),
\end{equation}
where $+$ applies when $E_{e} > 2m_{\chi}$ and $-$ applies when $E_{e} < 2m_{\chi}$.

Finally, by integration over $E_e$, we obtain the total number of 
detected events:
\begin{equation}
N =  \int_{E_{\mathrm{thr}}}^{\infty} 
\frac{dN}{dE_e}\,dE_e \label{total_N}, 
\end{equation}
where the SN burst duration is assumed to be $\Delta t_{\rm SN} = 10~\mathrm
s$, $E_{\mathrm{thr}}$ is the low energy threshold. A more general equation would also include the detection efficiency, which requires dedicated detector simulation. Here we take it as $100\%$ for simplicity, which suffices for the purpose of this work.

In our calculation, an experiment's sensitivity depends on the low-energy threshold $E_{\mathrm{thr}}$ and the number of target electrons $N_{\mathrm{T}}$. We analyze four representative experiments originally designed for neutrino or dark matter detection, XENONnT, JUNO, DUNE and Hyper-Kamiokande (Hyper-K), with specifics summarized in Table~\ref{tab:detectors}. \\

\begin{table}[t]
\centering
\caption{Detector specifics used in this analysis. 
$E_{\rm thr}$ denotes the detection threshold corresponding to minimum electron recoil energy; 
for JUNO and Hyper-K it corresponds to electron-equivalent 
visible energy. $N_{\mathrm{T}}$ is the number of target electrons 
contained in the listed mass. The XENONnT entry uses the 
full active LXe mass. The Hyper-K entry adopts a 30~kton effective fiducial; 
the full design value is 187~kton. The DUNE entry refers to the far detector and adopts a 40 kton fiducial mass; the full active LAr mass is $\sim$70 kton.}
\label{tab:detectors}
\begin{tabular}{lcccc}
\hline\hline
Detector & Target & Mass & $E_{\rm thr}$ & $N_{\mathrm{T}}$ \\
\hline
XENONnT~\cite{XENON:2023cxc}   & LXe    & 5.9~ton  & 1~keV    & $1.5\times10^{30}$ \\
JUNO~\cite{JUNO:2021vlw}       & LS     & 20~kton  & 0.2~MeV  & $6.4\times10^{33}$ \\
DUNE~\cite{DUNE:2020ypp}       & LAr    & 40~kton  & 5~MeV    & $1.1\times10^{34}$ \\
Hyper-K~\cite{HyperK:2018ofw}  & H$_2$O & 30~kton  & 3.5~MeV  & $1.0\times10^{34}$ \\
\hline\hline
\end{tabular}
\end{table}

\vspace{6pt}
\textbf{\textit{Background and Its Reduction.}}
The primary irreducible backgrounds to our mCP signal all arise from neutrinos: solar, atmospheric, and SN neutrino fluxes. Both neutrino-electron scattering and charged current neutrino-nucleus scattering in which the scattered nucleus is not detected, can fake our signal. 
However, as shown in Fig.~\ref{fig:time_profile}, the time profile of the mCP signal is a powerful way to discriminate between the approximately time-independent solar and atmospheric neutrino backgrounds and the prompt SN neutrino burst. Thus, one can optimize the timing cuts as a function of mCP mass, in order to minimize the contribution from these backgrounds. 
Angular cuts on the events would further reduce solar and atmospheric neutrino backgrounds when directional information is available. The supernova direction may be supplied by electromagnetic observations or by the neutrino burst itself; in particular, DUNE has dedicated pointing capabilities for Galactic supernova neutrinos~\cite{DUNE:2024ptd}. Furthermore, in a given time window, the electron recoil energy spectrum from mCP scattering (example given in Fig.~\ref{fig:recoil_spectrum} in Appendix~\ref{app:event_counts}) differs from that of neutrinos, which serves as an additional discriminator; related uses of spectral information in neutrino-detector searches for weakly coupled particles have been explored in Refs.~\cite{Cui:2022owf}. A detailed background mitigation strategy is beyond the scope of this work. For simplicity, the analysis adopts the reasonable assumption of zero background.

\section{Results}\label{sec:results}
\begin{figure*}[htbp]
    \centering
    \includegraphics[width=\textwidth]{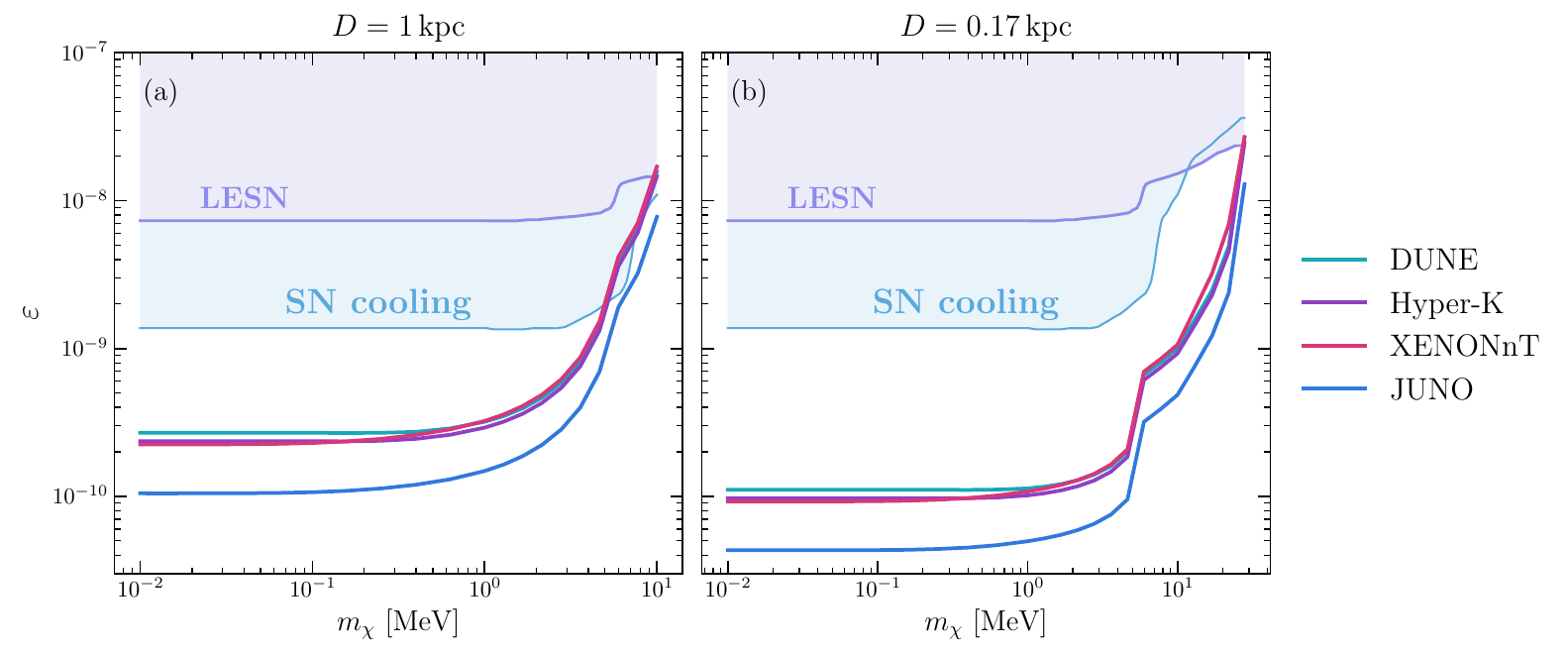}
    \caption{Projected discovery reach in the $(m_\chi,\varepsilon)$ plane for XENONnT, JUNO, DUNE, and Hyper-K, defined as 10 mCP-induced electron-recoil events within one year after the neutrino burst under the zero-background assumption. Panel~(a) assumes a supernova at $D=1~\mathrm{kpc}$, while panel~(b) assumes $D=0.17~\mathrm{kpc}$, corresponding to Betelgeuse~\cite{Joyce:2020vbx}. Both panels include production from plasmon decay and $e^+e^-$ annihilation. The blue shaded region shows the SN~1987A cooling bound and the purple shaded region shows the LESN exclusion region from Ref.~\cite{Li:2024pcp} (extrapolated to low masses).
    }
\label{fig:sensitivity_PD}
\end{figure*}

The core results of this work concern the mCP signal at the four terrestrial detectors listed in Table~\ref{tab:detectors}, following a future Galactic core-collapse supernova. The total number of events within the search window determines the sensitivity of the signal. Here, a discovery is defined as observing 10 or more mCP-induced events within the first year following the neutrino burst. 

The resulting sensitivity curves can easily be rescaled for a different choice of signal events, according to $N \propto \varepsilon^{4}$. This relation also implies that the projected reach in $\varepsilon$ have a weak sensitivity to the target number of signal events.

Fig.~\ref{fig:sensitivity_PD} shows the projected discovery reach under the zero-background assumption. Each curve gives the value of $\varepsilon$ for which 10 mCP-induced events are expected within one year after the neutrino burst. The points above a given curve therefore yield a larger signal sample, while those below it fall short of this benchmark. We use this 10 event criterion as a simple measure of sensitivity. For a benchmark distance of $D = 1$~kpc and a coupling of $\varepsilon = 10^{-9}$, all four detectors clear this discovery threshold for sub-MeV to MeV-scale masses. This reach extends below both the SN1987A cooling bound and the LESN bound by up to an order of magnitude in $\varepsilon$. JUNO delivers the strongest reach, attaining $\varepsilon_{\min} \sim 10^{-10}$ near the peak-sensitivity mass, while Hyper-K, DUNE, and XENONnT follow closely within a factor of a few. A comprehensive breakdown for individual detectors is provided in App.~\ref{app:event_counts}.

The overall shape of the sensitivity curve is governed by the interplay between production kinematics and the ToF selection. Plasmon decay sets the low-mass floor and shuts off near $m_\chi \sim 6$~MeV because the in-medium photon no longer has sufficient timelike invariant mass to produce a $\chi\bar{\chi}$ pair. By contrast, the $e^+e^-$ annihilation channel varies only mildly with $m_\chi$ in the low-mass regime, since the relevant center-of-mass energies are controlled mainly by the thermal and in-medium electron scales. It extends the reach after plasmon decay shuts off and dominates above $m_\chi \sim 57$~MeV. The sharp rise toward larger masses is driven primarily by the ToF cutoff: according to Eq.~\eqref{eq:time_delay}, mCPs reaching Earth within the one-year window must satisfy the energy condition $E_\chi \gtrsim m_\chi \sqrt{D/(2\Delta t_{\rm obs})}$. Note that proton bremsstrahlung, which is omitted here, would further improve the reach across the intermediate-mass band, rendering these projections inherently conservative.

A closer SN source would also extend the sensitivity reach. Panel~(b) of Fig.~\ref{fig:sensitivity_PD} illustrates this effect for $D = 0.17$~kpc, the distance to the nearby red supergiant Betelgeuse, following the $\varepsilon \propto \sqrt{D}$ scaling outlined in App.~\ref{app:past_SNe}.

If an excess of events is observed, two additional features can help confirm its mCP origin. First, the time profile shown in Fig.~\ref{fig:time_profile} distinguishes the delayed mCPs from the prompt neutrino burst. Second, the electron recoil spectrum within a fixed arrival time window (App.~\ref{app:event_counts}) provides an additional handle to mitigate backgrounds while containing information on the mCP mass.

\section{Conclusions and Outlook}\label{sec:conclusions}

While traditional supernova constraints on mCPs rely on their impact on SN dynamics, this work utilizes the escaping mCP flux that enables direct terrestrial detection. Production within the core proceeds through processes such as plasmon decay and $e^+e^-$ annihilation. The mass-dependent ToF delay then separates the arriving mCP signal from the SN neutrino burst, while the distinct electron recoil spectrum provides an additional signature. With XENONnT, JUNO, DUNE, and Hyper-K serving as target facilities, a future Galactic core-collapse supernova would yield a discovery reach that extends below the SN1987A cooling and LESN bounds across the sub-MeV to MeV mass range. 

This SN-based new search strategy for mCPs offers an important complement to laboratory searches~\cite{Prinz:1998ua, ArgoNeuT:2019ckq, Ball:2020dnx}. While accelerator and collider experiments probe the higher-mass and stronger-coupling regimes of the mCP parameter space, a SN explosion sheds light on the low-mass range. Remarkably, this sensitivity is achievable without the need for new instrumentation. Because the analysis operates as a delayed-arrival search on data already recorded under SNEWS-coordinated~\cite{Kara:2024xug} burst triggers, it leverages existing experimental infrastructure at zero additional hardware cost. Furthermore, depending on the detector, the maximum detection distance extends from $\sim 18$~kpc up to $\sim 83$~kpc (JUNO), with further details provided in App.~\ref{app:past_SNe}.

The next Galactic supernova may therefore serve not only as a pristine laboratory for neutrino physics and stellar dynamics, but also as a uniquely sensitive instrument for discovering dark sector particles such as mCPs.

\begin{acknowledgments}
 YC and FL are supported by the US Department of Energy under award number DE-SC0008541. IMS and XQ are supported by the U.S. Department of Energy under award number DE-SC0020262.
 YDT is supported by the Royal Society Dorothy Hodgkin Fellowship, and appreciates the generous start-up support from the University of Manchester and the University of Sheffield.
\end{acknowledgments}

\clearpage

\bibliography{bibliography}

@article{DUNE:2024ptd,
    author = "Abed Abud, Adam and others",
    collaboration = "DUNE",
    title = "{Supernova pointing capabilities of DUNE}",
    eprint = "2407.10339",
    archivePrefix = "arXiv",
    primaryClass = "hep-ex",
    reportNumber = "FERMILAB-PUB-24-0319-LBNF",
    doi = "10.1103/PhysRevD.111.092006",
    journal = "Phys. Rev. D",
    volume = "111",
    number = "9",
    pages = "092006",
    year = "2025"
}

@article{Tsai:2024wdh,
    author = "Tsai, Yu-Dai and Hwang, Insung and Schmitz, Ryan and Citron, Matthew and Gunthoti, Kranti and Steenis, Jacob and Jeong, Hoyong and Moon, Hyunki and Yoo, Jae Hyeok and Liu, Ming Xiong",
    title = "{Proposal for a dedicated search for millicharged and fractionally charged particles at LANL}",
    eprint = "2407.07142",
    archivePrefix = "arXiv",
    primaryClass = "hep-ph",
    reportNumber = "LA-UR-24-27441, FERMILAB-PUB-24-0357-T-V",
    doi = "10.1103/4bdh-sd5n",
    journal = "Phys. Rev. D",
    volume = "113",
    number = "1",
    pages = "015038",
    year = "2026"
}

@article{Gan:2023jbs,
    author = "Gan, Xucheng and Tsai, Yu-Dai",
    title = "{Cosmic millicharge background and reheating probes}",
    eprint = "2308.07951",
    archivePrefix = "arXiv",
    primaryClass = "hep-ph",
    reportNumber = "UCI-HEP-TR-2023-05, FERMILAB-PUB-23-428-T-V",
    doi = "10.1007/JHEP07(2025)094",
    journal = "JHEP",
    volume = "07",
    pages = "094",
    year = "2025"
}

@article{Iles:2024zka,
    author = "Iles, Ella and Heeba, Saniya and Schutz, Katelin",
    title = "{Dark Matter Direct Detection Experiments Are Sensitive to the Millicharged Background}",
    eprint = "2407.21096",
    archivePrefix = "arXiv",
    primaryClass = "hep-ph",
    doi = "10.1103/PhysRevLett.134.121002",
    journal = "Phys. Rev. Lett.",
    volume = "134",
    number = "12",
    pages = "121002",
    year = "2025"
}

@article{Boddy:2024vgt,
    author = "Boddy, Kimberly K. and Freese, Katherine and Montefalcone, Gabriele and Shams Es Haghi, Barmak",
    title = "{Minimal dark matter freeze-in with low reheating temperatures and implications for direct detection}",
    eprint = "2405.06226",
    archivePrefix = "arXiv",
    primaryClass = "hep-ph",
    reportNumber = "UTWI-16-2024, NORDITA-2024-016",
    doi = "10.1103/PhysRevD.111.063537",
    journal = "Phys. Rev. D",
    volume = "111",
    number = "6",
    pages = "063537",
    year = "2025"
}

@article{Dvorkin:2019zdi,
    author = "Dvorkin, Cora and Lin, Tongyan and Schutz, Katelin",
    title = "{Making dark matter out of light: freeze-in from plasma effects}",
    eprint = "1902.08623",
    archivePrefix = "arXiv",
    primaryClass = "hep-ph",
    doi = "10.1103/PhysRevD.99.115009",
    journal = "Phys. Rev. D",
    volume = "99",
    number = "11",
    pages = "115009",
    year = "2019",
    note = "[Erratum: Phys.Rev.D 105, 119901 (2022)]"
}

@article{Caputo:2022mah,
    author = "Caputo, Andrea and Janka, Hans-Thomas and Raffelt, Georg and Vitagliano, Edoardo",
    title = "{Low-Energy Supernovae Severely Constrain Radiative Particle Decays}",
    eprint = "2201.09890",
    archivePrefix = "arXiv",
    primaryClass = "astro-ph.HE",
    doi = "10.1103/PhysRevLett.128.221103",
    journal = "Phys. Rev. Lett.",
    volume = "128",
    number = "22",
    pages = "221103",
    year = "2022"
}

@article{Bailloeul:2025fde,
    author = "Bailloeul, Leo and Citron, Matthew and Cui, Yanou and Foroughi-Abari, Saeid and Hwang, Insung and Li, Fengyi and Tsai, Yu-Dai and Liu, Ming Xiong and Gunthoti, Kranti and Yoo, Jae Hyeok",
    title = "{Dedicated Searches for Millicharged Particles at Intensity-Frontier Facilities: SpinQuest and SHiP}",
    eprint = "2512.11027",
    archivePrefix = "arXiv",
    primaryClass = "hep-ph",
    month = "12",
    year = "2025"
}

@article{Dirac:1931kp,
    author = "Dirac, P. A. M.",
    title = "{Quantised singularities in the electromagnetic field}",
    journal = "Proc. Roy. Soc. Lond. A",
    volume = "133",
    pages = "60--72",
    year = "1931"
}

@article{Wen:1985qj,
    author = "Wen, Xiao-Gang and Witten, Edward",
    title = "{Electric and Magnetic Charges in Superstring Models}",
    reportNumber = "Print-85-0468 (PRINCETON)",
    doi = "10.1016/0550-3213(85)90592-9",
    journal = "Nucl. Phys. B",
    volume = "261",
    pages = "651--677",
    year = "1985"
}

@article{Pati:1973uk,
      author         = "Pati, Jogesh C. and Salam, Abdus",
      title          = "{Unified Lepton-Hadron Symmetry and a Gauge Theory of the
                        Basic Interactions}",
      journal        = "Phys. Rev.",
      volume         = "D8",
      year           = "1973",
      pages          = "1240-1251",
      doi            = "10.1103/PhysRevD.8.1240",
      reportNumber   = "IC-73-41-INT-REP",
      SLACcitation   = "%%CITATION = PHRVA,D8,1240;%%"
}

@article{Lee:2026djo,
    author = "Lee, Junseok and Takahashi, Fuminobu and Tsai, Yu-Dai",
    title = "{Number Theory in Quantum Physics: Minicharged Particles and the Prouhet-Tarry-Escott Problem}",
    eprint = "2603.12320",
    archivePrefix = "arXiv",
    primaryClass = "hep-ph",
    reportNumber = "TU-1300",
    month = "3",
    year = "2026"
}

@article{Feldman:2007wj,
    author = "Feldman, Daniel and Liu, Zuowei and Nath, Pran",
    title = "{The Stueckelberg Z-prime Extension with Kinetic Mixing and Milli-Charged Dark Matter From the Hidden Sector}",
    eprint = "hep-ph/0702123",
    archivePrefix = "arXiv",
    doi = "10.1103/PhysRevD.75.115001",
    journal = "Phys. Rev. D",
    volume = "75",
    pages = "115001",
    year = "2007"
}

@article{Cheung:2007ut,
    author = "Cheung, Kingman and Yuan, Tzu-Chiang",
    title = "{Hidden fermion as milli-charged dark matter in Stueckelberg Z- prime model}",
    eprint = "hep-ph/0701107",
    archivePrefix = "arXiv",
    doi = "10.1088/1126-6708/2007/03/120",
    journal = "JHEP",
    volume = "03",
    pages = "120",
    year = "2007"
}

@article{Kelly:2018brz,
    author = "Kelly, Kevin J. and Tsai, Yu-Dai",
    title = "{Proton fixed-target scintillation experiment to search for millicharged dark matter}",
    eprint = "1812.03998",
    archivePrefix = "arXiv",
    primaryClass = "hep-ph",
    reportNumber = "FERMILAB-PUB-18-668-A-PPD-T",
    doi = "10.1103/PhysRevD.100.015043",
    journal = "Phys. Rev. D",
    volume = "100",
    number = "1",
    pages = "015043",
    year = "2019"
}

@article{Ahlers:2007qf,
    author = "Ahlers, M. and Gies, H. and Jaeckel, J. and Redondo, J. and Ringwald, A.",
    title = "{Laser experiments explore the hidden sector}",
    eprint = "0711.4991",
    archivePrefix = "arXiv",
    primaryClass = "hep-ph",
    reportNumber = "DESY-07-207, OUTP-0715P, IPPP-07-93, DCPT-07-186",
    doi = "10.1103/PhysRevD.77.095001",
    journal = "Phys. Rev. D",
    volume = "77",
    pages = "095001",
    year = "2008"
}

@article{ArguellesDelgado:2021lek,
    author = {Arg{\"u}elles Delgado, Carlos Alberto and Kelly, Kevin James and Mu{\~n}oz Albornoz, V{\'\i}ctor},
    title = "{Millicharged particles from the heavens: single- and multiple-scattering signatures}",
    eprint = "2104.13924",
    archivePrefix = "arXiv",
    primaryClass = "hep-ph",
    reportNumber = "FERMILAB-PUB-21-214-T",
    doi = "10.1007/JHEP11(2021)099",
    journal = "JHEP",
    volume = "11",
    pages = "099",
    year = "2021"
}

@article{Plestid:2020kdm,
    author = "Plestid, Ryan and Takhistov, Volodymyr and Tsai, Yu-Dai and Bringmann, Torsten and Kusenko, Alexander and Pospelov, Maxim",
    title = "{New Constraints on Millicharged Particles from Cosmic-ray Production}",
    eprint = "2002.11732",
    archivePrefix = "arXiv",
    primaryClass = "hep-ph",
    reportNumber = "FERMILAB-PUB-20-044-A-T, INT-PUB-20-004, IPMU20-0015",
    doi = "10.1103/PhysRevD.102.115032",
    journal = "Phys. Rev. D",
    volume = "102",
    pages = "115032",
    year = "2020"
}

@article{Harnik:2020ugb,
    author = "Harnik, Roni and Plestid, Ryan and Pospelov, Maxim and Ramani, Harikrishnan",
    title = "{Millicharged cosmic rays and low recoil detectors}",
    eprint = "2010.11190",
    archivePrefix = "arXiv",
    primaryClass = "hep-ph",
    reportNumber = "FERMILAB-PUB-20-523-T",
    doi = "10.1103/PhysRevD.103.075029",
    journal = "Phys. Rev. D",
    volume = "103",
    number = "7",
    pages = "075029",
    year = "2021"
}

@article{Berlin:2025btf,
    author = "Berlin, Asher and Bogorad, Zachary and Graham, Peter W. and Ramani, Harikrishnan",
    title = "{Electric Accumulation of Millicharged Particles}",
    eprint = "2510.25834",
    archivePrefix = "arXiv",
    primaryClass = "hep-ph",
    reportNumber = "FERMILAB-PUB-25-0622-SQMS-T",
    month = "10",
    year = "2025"
}

@article{Berlin:2025hjs,
    author = "Berlin, Asher and Bogorad, Zachary and Graham, Peter W. and Ramani, Harikrishnan",
    title = "{Cavendish Tests of Millicharged Particles}",
    eprint = "2510.25825",
    archivePrefix = "arXiv",
    primaryClass = "hep-ph",
    reportNumber = "FERMILAB-PUB-25-0623-SQMS-T",
    month = "10",
    year = "2025"
}

@article{Wu:2024iqm,
    author = "Wu, Han and Hardy, Edward and Song, Ningqiang",
    title = "{Searching for heavy millicharged particles from the atmosphere}",
    eprint = "2406.01668",
    archivePrefix = "arXiv",
    primaryClass = "hep-ph",
    doi = "10.1103/PhysRevD.110.115037",
    journal = "Phys. Rev. D",
    volume = "110",
    number = "11",
    pages = "115037",
    year = "2024"
}

@article{Berlin:2021kcm,
    author = "Berlin, Asher and Schutz, Katelin",
    title = "{Helioscope for gravitationally bound millicharged particles}",
    eprint = "2111.01796",
    archivePrefix = "arXiv",
    primaryClass = "hep-ph",
    reportNumber = "MIT-CTP/5358, FERMILAB-PUB-21-626-T",
    doi = "10.1103/PhysRevD.105.095012",
    journal = "Phys. Rev. D",
    volume = "105",
    number = "9",
    pages = "095012",
    year = "2022"
}

@article{Davidson:1993sj,
    author = "Davidson, Sacha and Peskin, Michael E.",
    title = "{Astrophysical bounds on millicharged particles in models with a paraphoton}",
    eprint = "hep-ph/9310288",
    archivePrefix = "arXiv",
    reportNumber = "SLAC-PUB-6360, CFPA-93-TH-31",
    doi = "10.1103/PhysRevD.49.2114",
    journal = "Phys. Rev. D",
    volume = "49",
    pages = "2114--2117",
    year = "1994"
}

@article{Berlin:2024dwg,
    author = "Berlin, Asher and Harnik, Roni and Li, Ying-Ying and Xu, Bin",
    title = "{Millicharged Condensates on Earth}",
    eprint = "2404.16094",
    archivePrefix = "arXiv",
    primaryClass = "hep-ph",
    reportNumber = "FERMILAB-PUB-24-0002-SQMS-T, USTC-ICTS/PCFT-24-08",
    month = "4",
    year = "2024"
}

@article{Cappiello:2019qsw,
    author = "Cappiello, Christopher V. and Beacom, John F.",
    title = "{Strong New Limits on Light Dark Matter from Neutrino Experiments}",
    eprint = "1906.11283",
    archivePrefix = "arXiv",
    primaryClass = "hep-ph",
    doi = "10.1103/PhysRevD.104.069901",
    journal = "Phys. Rev. D",
    volume = "100",
    number = "10",
    pages = "103011",
    year = "2019",
    note = "[Erratum: Phys.Rev.D 104, 069901 (2021)]"
}

@article{Magill:2018tbb,
    author = "Magill, Gabriel and Plestid, Ryan and Pospelov, Maxim and Tsai, Yu-Dai",
    title = "{Millicharged particles in neutrino experiments}",
    eprint = "1806.03310",
    archivePrefix = "arXiv",
    primaryClass = "hep-ph",
    reportNumber = "FERMILAB-PUB-18-631-A",
    doi = "10.1103/PhysRevLett.122.071801",
    journal = "Phys. Rev. Lett.",
    volume = "122",
    number = "7",
    pages = "071801",
    year = "2019"
}

@article{Chang:2018rso,
    author = "Chang, Jae Hyeok and Essig, Rouven and McDermott, Samuel D.",
    title = "{Supernova 1987A Constraints on Sub-GeV Dark Sectors, Millicharged Particles, the QCD Axion, and an Axion-like Particle}",
    eprint = "1803.00993",
    archivePrefix = "arXiv",
    primaryClass = "hep-ph",
    reportNumber = "YITP-SB-18-01, FERMILAB-PUB-17-432-T",
    doi = "10.1007/JHEP09(2018)051",
    journal = "JHEP",
    volume = "09",
    pages = "051",
    year = "2018"
}

@article{Fiorillo:2024upk,
    author = "Fiorillo, Damiano F. G. and Vitagliano, Edoardo",
    title = "{Self-Interacting Dark Sectors in Supernovae Can Behave as a Relativistic Fluid}",
    eprint = "2404.07714",
    archivePrefix = "arXiv",
    primaryClass = "hep-ph",
    doi = "10.1103/PhysRevLett.133.251004",
    journal = "Phys. Rev. Lett.",
    volume = "133",
    number = "25",
    pages = "251004",
    year = "2024"
}

@article{Davidson:2000hf,
    author = "Davidson, Sacha and Hannestad, Steen and Raffelt, Georg",
    title = "{Updated bounds on millicharged particles}",
    eprint = "hep-ph/0001179",
    archivePrefix = "arXiv",
    reportNumber = "CERN-TH-99-384",
    doi = "10.1088/1126-6708/2000/05/003",
    journal = "JHEP",
    volume = "05",
    pages = "003",
    year = "2000"
}

@article{Cui:2022owf,
    author = "Cui, Yanou and Kuo, Jui-Lin and Pradler, Josef and Tsai, Yu-Dai",
    title = "{Shining light on cosmogenic axions with neutrino experiments}",
    eprint = "2207.13107",
    archivePrefix = "arXiv",
    primaryClass = "hep-ph",
    reportNumber = "UCI-HEP-TR-2022-12, FERMILAB-PUB-22-883-T",
    doi = "10.1103/PhysRevD.106.115024",
    journal = "Phys. Rev. D",
    volume = "106",
    number = "11",
    pages = "115024",
    year = "2022"
}

@article{SENSEI:2023gie,
    author = "Barak, Liron and others",
    collaboration = "SENSEI",
    title = "{Search by the SENSEI Experiment for Millicharged Particles Produced in the NuMI Beam}",
    eprint = "2305.04964",
    archivePrefix = "arXiv",
    primaryClass = "hep-ex",
    reportNumber = "CALT-TH-2023-011, YITP-SB-2023-07, FERMILAB-PUB-23-222-PPD",
    doi = "10.1103/PhysRevLett.133.071801",
    journal = "Phys. Rev. Lett.",
    volume = "133",
    number = "7",
    pages = "071801",
    year = "2024"
}

@article{Oscura:2023qch,
    author = "Perez, Santiago and others",
    collaboration = "Oscura",
    title = "{Searching for millicharged particles with 1 kg of Skipper-CCDs using the NuMI beam at Fermilab}",
    eprint = "2304.08625",
    archivePrefix = "arXiv",
    primaryClass = "hep-ex",
    reportNumber = "FERMILAB-PUB-23-200-PPD-T",
    doi = "10.1007/JHEP02(2024)072",
    journal = "JHEP",
    volume = "02",
    pages = "072",
    year = "2024"
}

@article{Foroughi-Abari:2020qar,
    author = "Foroughi-Abari, Saeid and Kling, Felix and Tsai, Yu-Dai",
    title = "{Looking forward to millicharged dark sectors at the LHC}",
    eprint = "2010.07941",
    archivePrefix = "arXiv",
    primaryClass = "hep-ph",
    reportNumber = "FERMILAB-PUB-20-477-AE-PPD-T",
    doi = "10.1103/PhysRevD.104.035014",
    journal = "Phys. Rev. D",
    volume = "104",
    number = "3",
    pages = "035014",
    year = "2021"
}

@article{Berlin:2018bsc,
    author = "Berlin, Asher and Blinov, Nikita and Krnjaic, Gordan and Schuster, Philip and Toro, Natalia",
    title = "{Dark Matter, Millicharges, Axion and Scalar Particles, Gauge Bosons, and Other New Physics with LDMX}",
    eprint = "1807.01730",
    archivePrefix = "arXiv",
    primaryClass = "hep-ph",
    reportNumber = "FERMILAB-PUB-18-310-A, SLAC-PUB-17297",
    doi = "10.1103/PhysRevD.99.075001",
    journal = "Phys. Rev. D",
    volume = "99",
    number = "7",
    pages = "075001",
    year = "2019"
}

@article{Essig:2024dpa,
    author = "Essig, Rouven and Li, Peiran and Liu, Zhen and McDuffie, Megan and Plestid, Ryan and Xu, Hailin",
    title = "{Probing millicharged particles at an electron beam dump with ultralow-threshold sensors}",
    eprint = "2412.09652",
    archivePrefix = "arXiv",
    primaryClass = "hep-ph",
    doi = "10.1007/JHEP04(2025)057",
    journal = "JHEP",
    volume = "04",
    pages = "057",
    year = "2025"
}

@article{DellaValle:2014xoa,
    author = "Della Valle, F. and Milotti, E. and Ejlli, A. and Messineo, G. and Piemontese, L. and Zavattini, G. and Gastaldi, U. and Pengo, R. and Ruoso, G.",
    title = "{First results from the new PVLAS apparatus: A new limit on vacuum magnetic birefringence}",
    eprint = "1406.6518",
    archivePrefix = "arXiv",
    primaryClass = "quant-ph",
    doi = "10.1103/PhysRevD.90.092003",
    journal = "Phys. Rev. D",
    volume = "90",
    number = "9",
    pages = "092003",
    year = "2014"
}

@article{Holdom:1985ag,
    author = "Holdom, Bob",
    title = "{Two U(1)'s and Epsilon Charge Shifts}",
    reportNumber = "UTPT-85-30",
    doi = "10.1016/0370-2693(86)91377-8",
    journal = "Phys. Lett. B",
    volume = "166",
    pages = "196--198",
    year = "1986"
}

@article{Kara:2024xug,
    author = "Kara, M. and others",
    title = "{The SNEWS~2.0 alert software for the coincident detection of neutrinos from core-collapse supernovae}",
    eprint = "2406.17743",
    archivePrefix = "arXiv",
    primaryClass = "astro-ph.IM",
    doi = "10.1088/1748-0221/19/10/P10017",
    journal = "JINST",
    volume = "19",
    number = "10",
    pages = "P10017",
    year = "2024"
}

@misc{zcoupling,
note         = {The coupling to $Z$ is suppressed and the effect can be neglected for the processes we consider.}
}

@article{Li:2024pcp,
    author = "Li, Changqian and Liu, Zuowei and Lu, Wenxi and Ye, Zicheng",
    title = "{Low-energy supernova constraints on millicharged particles}",
    eprint = "2408.04953",
    archivePrefix = "arXiv",
    primaryClass = "hep-ph",
    doi = "10.1007/JHEP07(2025)116",
    journal = "JHEP",
    volume = "07",
    pages = "116",
    year = "2025"
}

@article{Holdom:1986eq,
    author = "Holdom, Bob",
    title = "{Searching for epsilon charges and a new U(1)}",
    doi = "10.1016/0370-2693(86)90470-3",
    journal = "Phys. Lett. B",
    volume = "178",
    pages = "65--70",
    year = "1986"
}

@article{XENON:2023cxc,
    author = "Aprile, E. and others",
    collaboration = "XENON",
    title = "{First Dark Matter Search with Nuclear Recoils from the XENONnT Experiment}",
    eprint = "2303.14729",
    archivePrefix = "arXiv",
    primaryClass = "hep-ex",
    doi = "10.1103/PhysRevLett.131.041003",
    journal = "Phys. Rev. Lett.",
    volume = "131",
    pages = "041003",
    year = "2023"
}

@article{JUNO:2021vlw,
    author = "Abusleme, Angel and others",
    collaboration = "JUNO",
    title = "{JUNO Physics and Detector}",
    eprint = "2104.02565",
    archivePrefix = "arXiv",
    primaryClass = "hep-ex",
    doi = "10.1016/j.ppnp.2022.103927",
    journal = "Prog. Part. Nucl. Phys.",
    volume = "123",
    pages = "103927",
    year = "2022"
}

@article{DUNE:2020ypp,
    author = "Abi, B. and others",
    collaboration = "DUNE",
    title = "{Prospects for beyond the standard model physics searches 
              at the Deep Underground Neutrino Experiment}",
    eprint = "2008.12769",
    archivePrefix = "arXiv",
    primaryClass = "hep-ex",
    doi = "10.1140/epjc/s10052-021-09007-w",
    journal = "Eur. Phys. J. C",
    volume = "81",
    pages = "322",
    year = "2021"
}

@article{HyperK:2018ofw,
    author = "Abe, K. and others",
    collaboration = "Hyper-Kamiokande",
    title = "{Hyper-Kamiokande Design Report}",
    eprint = "1805.04163",
    archivePrefix = "arXiv",
    primaryClass = "physics.ins-det",
    month = "5",
    year = "2018"
}

@article{Braaten:1993jw,
    author = "Braaten, Eric and Segel, Daniel",
    title = "{Neutrino energy loss from the plasma process 
              at all temperatures and densities}",
    eprint = "hep-ph/9302213",
    archivePrefix = "arXiv",
    doi = "10.1103/PhysRevD.48.1478",
    journal = "Phys. Rev. D",
    volume = "48",
    pages = "1478--1491",
    year = "1993"
}

@book{Raffelt:1996wa,
    author = "Raffelt, Georg G.",
    title = "{Stars as Laboratories for Fundamental Physics}",
    publisher = "University of Chicago Press",
    year = "1996"
}

@article{Chu:2019rok,
    author = "Chu, Xiaoyong and Kuo, Jui-Lin and Pradler, Josef and Semmelrock, Lukas",
    title = "{Stellar probes of dark sector-photon interactions}",
    eprint = "1908.00553",
    archivePrefix = "arXiv",
    primaryClass = "hep-ph",
    doi = "10.1103/PhysRevD.100.083002",
    journal = "Phys. Rev. D",
    volume = "100",
    pages = "083002",
    year = "2019"
}

@article{Hardy:2016kme,
    author = "Hardy, Edward and Lasenby, Robert",
    title = "{Stellar cooling bounds on new light particles: plasma mixing effects}",
    eprint = "1611.05852",
    archivePrefix = "arXiv",
    primaryClass = "hep-ph",
    doi = "10.1007/JHEP02(2017)033",
    journal = "JHEP",
    volume = "02",
    pages = "033",
    year = "2017"
}

@article{SNEWS:2021ezc,
    author = "Al Kharusi, S. and others",
    title = "{SNEWS 2.0: A Next-Generation SuperNova Early Warning System for Multi-messenger Astronomy}",
    journal = "New J. Phys.",
    volume = "23",
    pages = "031201",
    year = "2021",
    eprint = "2011.00035",
    archivePrefix = "arXiv",
    primaryClass = "astro-ph.HE"
}

@article{Essig:2013lka,
    author = "Essig, Rouven and others",
    title = "{Dark Sectors and New, Light, Weakly-Coupled Particles}",
    eprint = "1311.0029",
    archivePrefix = "arXiv",
    primaryClass = "hep-ph",
    month = "10",
    year = "2013"
}

@article{Battaglieri:2017aum,
    author = "Battaglieri, Marco and others",
    title = "{US Cosmic Visions: New Ideas in Dark Matter 2017: Community Report}",
    eprint = "1707.04591",
    archivePrefix = "arXiv",
    primaryClass = "hep-ph",
    reportNumber = "FERMILAB-CONF-17-282-AE-PPD-T",
    month = "7",
    year = "2017"
}

@article{Gori:2022vri,
    author = "Gori, Stefania and others",
    title = "{Dark Sector Physics at High-Intensity Experiments}",
    eprint = "2209.04671",
    archivePrefix = "arXiv",
    primaryClass = "hep-ph",
    month = "9",
    year = "2022"
}

@article{Joyce:2020vbx,
    author = "Joyce, Meridith and Leung, Shing-Chi and Moln\'ar, L\'aszl\'o and Ireland, Michael and Kobayashi, Chiaki and Nomoto, Ken'ichi",
    title = "{Standing on the Shoulders of Giants: New Mass and Distance Estimates for Betelgeuse through Combined Evolutionary, Asteroseismic, and Hydrodynamic Simulations with MESA}",
    eprint = "2006.09837",
    archivePrefix = "arXiv",
    primaryClass = "astro-ph.SR",
    doi = "10.3847/1538-4357/abb8db",
    journal = "Astrophys. J.",
    volume = "902",
    pages = "63",
    year = "2020"
}

@article{Prinz:1998ua,
    author = "Prinz, A.~A. and others",
    title = "{Search for millicharged particles at SLAC}",
    eprint = "hep-ex/9804008",
    archivePrefix = "arXiv",
    doi = "10.1103/PhysRevLett.81.1175",
    journal = "Phys. Rev. Lett.",
    volume = "81",
    pages = "1175--1178",
    year = "1998"
}

@article{ArgoNeuT:2019ckq,
    author = "Acciarri, R. and others",
    collaboration = "ArgoNeuT",
    title = "{Improved Limits on Millicharged Particles Using the ArgoNeuT Experiment at Fermilab}",
    eprint = "1911.07996",
    archivePrefix = "arXiv",
    primaryClass = "hep-ex",
    doi = "10.1103/PhysRevLett.124.131801",
    journal = "Phys. Rev. Lett.",
    volume = "124",
    number = "13",
    pages = "131801",
    year = "2020"
}

@article{Ball:2020dnx,
    author = "Ball, A. and others",
    collaboration = "milliQan",
    title = "{Search for millicharged particles in proton-proton collisions at $\sqrt{s}=13$~TeV}",
    eprint = "2005.06518",
    archivePrefix = "arXiv",
    primaryClass = "hep-ex",
    doi = "10.1103/PhysRevD.102.032002",
    journal = "Phys. Rev. D",
    volume = "102",
    number = "3",
    pages = "032002",
    year = "2020"
}

\appendix
\onecolumngrid

\section{mCP Production Rate}
\label{app:prod_rate}

This appendix summarizes the mCP production rates from plasmon decay and $e^{+}e^{-}$ annihilation in the one-zone SN model used throughout this work. The plasmon-decay calculation follows the finite-temperature plasma formalism of Ref.~\cite{Braaten:1993jw}, as implemented for mCP production in Ref.~\cite{Li:2024pcp}. The $e^{+}e^{-}$ annihilation calculation follows Refs.~\cite{Li:2024pcp,Chu:2019rok}.\\

Let's first consider plasmon decay. An in-medium photon of polarization $a$ (transverse ($T$) or longitudinal ($L$))
decays to a $\chi\bar{\chi}$ pair with the rate
\begin{equation}
\Gamma_a = Z_a \,\frac{\varepsilon^2 \alpha \, K^2}{3\,\omega_a}\,
f\!\left(\frac{m_\chi^2}{K^2}\right) ,
\label{eq:plasmon_width}
\end{equation}
where $K^\mu = (\omega, \mathbf{k})$ denotes the in-medium photon
four-momentum, $\alpha = e^2/4\pi$,
$Z_a$ is the wave-function renormalization
factor~\cite{Braaten:1993jw},
and $f(x) = \sqrt{1-4x}\,(1+2x)$ encodes the two-body phase space.
Integrating over the thermal photon distribution yields
the differential mCP production rate per unit volume and energy:
\begin{equation}
\frac{d\Phi_a}{dE_\chi} = \frac{g_a}{2\pi^2}
\int_0^\infty dk\, k^2 \,
\frac{\Gamma_a}{e^{\omega_a/T_c} - 1}\,
g(E_\chi, m_\chi, K) \,,
\label{eq:plasmon_rate_app}
\end{equation}
where $g_T = 2$, $g_L = 1$, and
\begin{equation}
g(E_\chi, m_\chi, K) = 2\,
\frac{\Theta(E_\chi - E_\chi^-)\,\Theta(E_\chi^+ - E_\chi)}
{E_\chi^+ - E_\chi^-}
\label{eq:spectral_g}
\end{equation}
is a flat spectral weight over the kinematically allowed
mCP energies
$E_\chi^\pm = \bigl(\omega \pm k\sqrt{1-4m_\chi^2/K^2}\bigr)/2$.
The prefactor of 2 counts the $\chi$ and $\bar{\chi}$
from each decay.\\

In the relativistic regime ($T_c \gg m_e$, $\mu_e \gg m_e$), where $T_c = 30$~MeV is the core temperature and $\mu_e \simeq 167$~MeV is the electron chemical potential,
the plasma frequency
takes the form of~\cite{Braaten:1993jw}
\begin{equation}
\omega_p^2 = \frac{4\alpha}{3\pi}
\left(\mu_e^2 + \frac{\pi^2 T_c^2}{3}\right) ,
\label{eq:plasma_freq}
\end{equation}
which leads to $\omega_p \simeq 9.8$~MeV for the parameters in the one-zone model.
The dispersion relations $\omega_{T,L}(k)$
and residue factors $Z_{T,L}(k)$ for each polarization
follow the transcendental equations
in Refs.~\cite{Braaten:1993jw,Li:2024pcp}.\\

\begin{figure*}[htbp]
    \centering    \includegraphics[width=\linewidth]{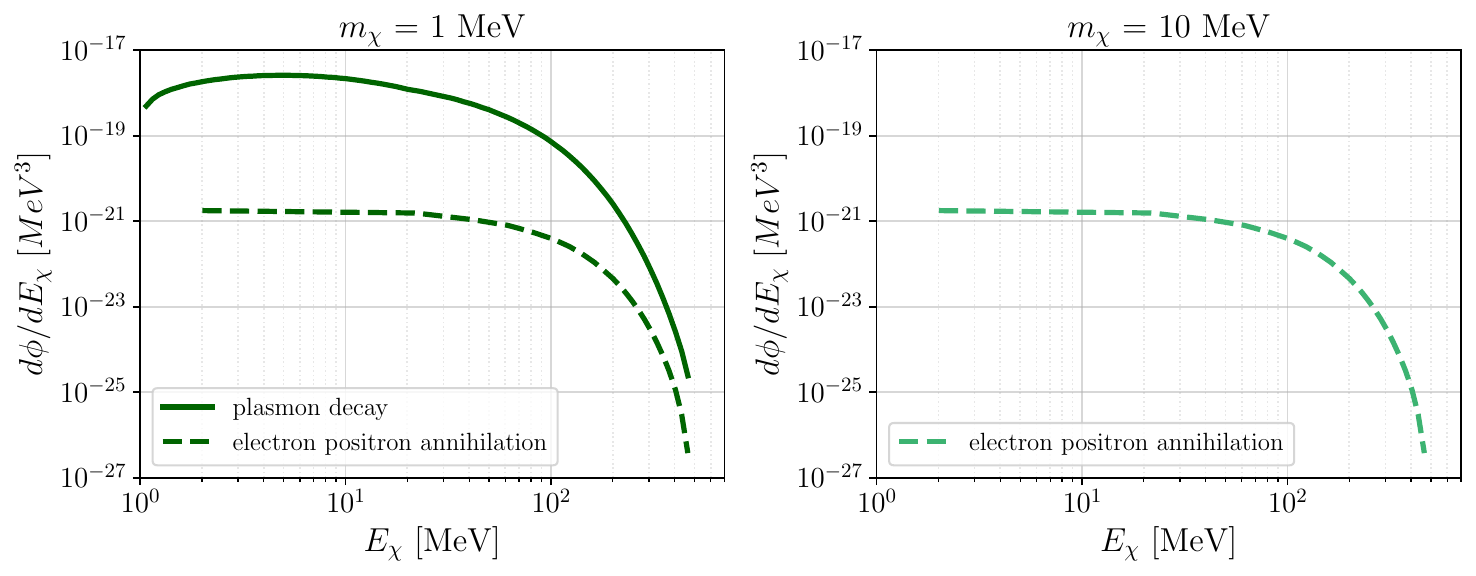}
    \caption{The left panel shows the mCP production rate in the SN core via the plasmon-decay channel (solid line) and the $e^{+}e^{-}$ annihilation channel (dashed line), with $m_{\chi} = 1~\mathrm{MeV}$. The right panel shows the corresponding result for $m_{\chi} = 10~\mathrm{MeV}$, where the plasmon-decay contribution is absent. The $e^{+}e^{-}$ annihilation spectra in the two panels are similar because, for these masses, the annihilation rate is controlled mainly by the thermal and in-medium electron scales rather than by $m_\chi$ itself.}
  \label{fig:prod_rate}
\end{figure*}

In Fig.~\ref{fig:prod_rate}, the production rate via the plasmon-decay channel at $m_{\chi}=1~\mathrm{MeV}$ is shown as the solid line in the left panel. This channel dominates at low mCP masses, but it is kinematically shut off once the in-medium photon cannot provide enough timelike invariant mass to produce a $\chi\bar{\chi}$ pair. This threshold appears explicitly in Eq.~\eqref{eq:plasmon_width} through the two-body phase-space factor
$f(m_\chi^2/K^2)=\sqrt{1-4m_\chi^2/K^2}\,(1+2m_\chi^2/K^2)$, which requires $K^2\geq 4m_\chi^2$. The same kinematic condition also appears in the endpoints $E_\chi^\pm$ defined below Eq.~\eqref{eq:spectral_g}. For the one-zone SN core parameters used here, the plasmon dispersion relations and residues give appreciable support only up to $m_\chi\simeq 6~\mathrm{MeV}$. Therefore the plasmon-decay contribution is absent for $m_\chi=10~\mathrm{MeV}$, as shown in the right panel of Fig.~\ref{fig:prod_rate}. \\

For the $e^+e^-$ annihilation channel,
the in-medium annihilation cross section
$e^+e^- \to \gamma^* \to \chi\bar{\chi}$ splits into
transverse and longitudinal channels,
$\sigma_{\rm ann} = \sigma_T + \sigma_L$,
with~\cite{Li:2024pcp}
\begin{equation}
\sigma_a = \frac{2\pi\varepsilon^2\alpha^2}{3\beta_e}\,
\frac{N_a \, K^2 \, f(m_\chi^2/K^2)}
{(K^2 - {\rm Re}\,\Pi_a)^2 + ({\rm Im}\,\Pi_a)^2} \,,
\label{eq:ann_xsec}
\end{equation}
where $\beta_e = \sqrt{1 - 4m_e^2/K^2}$, $\Pi_a$ is the
electromagnetic polarization tensor,
$N_a$ are kinematic prefactors defined in
Ref.~\cite{Li:2024pcp},
and $f(x)$ appears below Eq.~\eqref{eq:plasmon_width}.
The imaginary part of $\Pi_a$ shifts the cross section by no more
than ${\sim}\,2\%$~\cite{Li:2024pcp} and is dropped hereafter.
The in-medium electron mass
$m_e^{\rm eff} \simeq 9$~MeV is extracted from the
self-consistent gap equation in
Ref.~\cite{Hardy:2016kme}.\\

The differential production rate per unit volume per unit energy
for this channel is~\cite{Li:2024pcp,Chu:2019rok}
\begin{equation}
\frac{d\Phi_{\rm ann}}{dE_\chi} = \frac{1}{16\pi^4}
\int_{4m_{\rm th}^2}^{\infty} \!\! dK^2 \, K^2 \beta_e
\int_{\sqrt{K^2}}^{\infty} \!\! dE_+
\int_{-E_-^m}^{E_-^m} \!\! dE_- \;
f_1 f_2 \, \sigma_{\rm ann} \, g(E_\chi, m_\chi, K) \,,
\label{eq:ann_rate}
\end{equation}
where $E_\pm = E_1 \pm E_2$ are the sum and difference of the
initial $e^\pm$ energies, respectively, $f_{1,2}$ denote
Fermi--Dirac distributions evaluated at electron chemical potential
$\mu_e \simeq 167$~MeV,
$m_{\rm th} = \max(m_e^{\rm eff}, m_\chi)$,
$E_-^m = \sqrt{1 - 4m_e^2/K^2}\,\sqrt{E_+^2 - K^2}$,
and $g(E_\chi, m_\chi, K)$ is given in
Eq.~\eqref{eq:spectral_g}. The $m_\chi$ dependence of the annihilation rate enters through the phase-space factor
$f(m_\chi^2/K^2)$, the spectral endpoints in $g(E_\chi,m_\chi,K)$, and the lower integration limit
$m_{\rm th}=\max(m_e^{\rm eff},m_\chi)$. Therefore the rate is not exactly mass independent. However, for
$m_\chi$ below or comparable to the in-medium electron mass, $m_e^{\rm eff}\simeq 9~\mathrm{MeV}$, and below the typical thermally accessible center-of-mass energies in the SN core, these mass-dependent factors only mildly modify the integral. This explains why the $e^+e^-$ annihilation spectra shown as dashed lines in Fig.~\ref{fig:prod_rate} are nearly unchanged between $m_\chi=1~\mathrm{MeV}$ and $m_\chi=10~\mathrm{MeV}$. The mass dependence becomes more pronounced at larger $m_\chi$, where the final-state phase space shrinks and the threshold $K^2\geq 4m_\chi^2$ removes an increasing fraction of the thermally populated phase space.

\clearpage

\section{Electron-Recoil Spectra, Time Profiles, and Event Counts at $D=1~{\rm kpc}$}
\label{app:event_counts}

This appendix presents the integrated event counts, electron-recoil spectra, and arrival-time profiles for a Galactic core-collapse supernova at $D = 1$~kpc and $\varepsilon = 10^{-9}$, which serves as the benchmark throughout Sec.~\ref{sec:results}.

Table~\ref{tab:results} reports the total event count within the first year after the neutrino burst for four representative mCP masses. JUNO yields the largest sample, reaching $\mathcal{O}(10^4)$ to $\mathcal{O}(10^5)$ events for $m_\chi \lesssim 1$~MeV. Hyper-K, DUNE, and XENONnT record $\mathcal{O}(10^3)$ events in the same mass range, where the keV-scale recoil threshold of XENONnT partially compensates for its smaller target mass. The sharp suppression at $m_\chi = 10$~MeV is due to the absence of the plasmon decay channel and the ToF cutoff discussed in Sec.~\ref{sec:results}; it should not be interpreted as a sudden suppression of the $e^+e^-$ annihilation production rate itself, which remains only slightly mass dependent around this mass range.

\begin{table}[h]
\centering
\caption{Total event count within the first year after the neutrino burst at $D = 1$~kpc and $\varepsilon = 10^{-9}$, for four representative mCP masses (in MeV).}
\label{tab:results}
\begin{tabular}{lcccc}
\hline\hline
Detector & $m_{\chi} = 0.01$ & $m_{\chi} = 0.1$ &
$m_{\chi} = 1$ & $m_{\chi} = 10$ \\
\hline
XENONnT & $3.9\times10^3$  & $3.6\times 10^3$
    & $9.1\times10^2$  & $1.2\times10^{-4}$ \\
JUNO    & $8.3\times 10^4$ & $7.7\times 10^4$
    & $2.0\times 10^4$ & $2.7\times10^{-3}$ \\
DUNE    & $1.9\times10^3$  & $1.9\times 10^3$
    & $9.6\times 10^2$  & $1.6\times10^{-4}$ \\
Hyper-K & $3.2\times10^3$  & $3.2\times 10^3$
    & $1.4\times 10^3$  & $2.2\times10^{-4}$ \\
\hline\hline
\end{tabular}
\end{table}

\begin{figure*}[htbp]
    \centering    \includegraphics[width=0.8\textwidth]{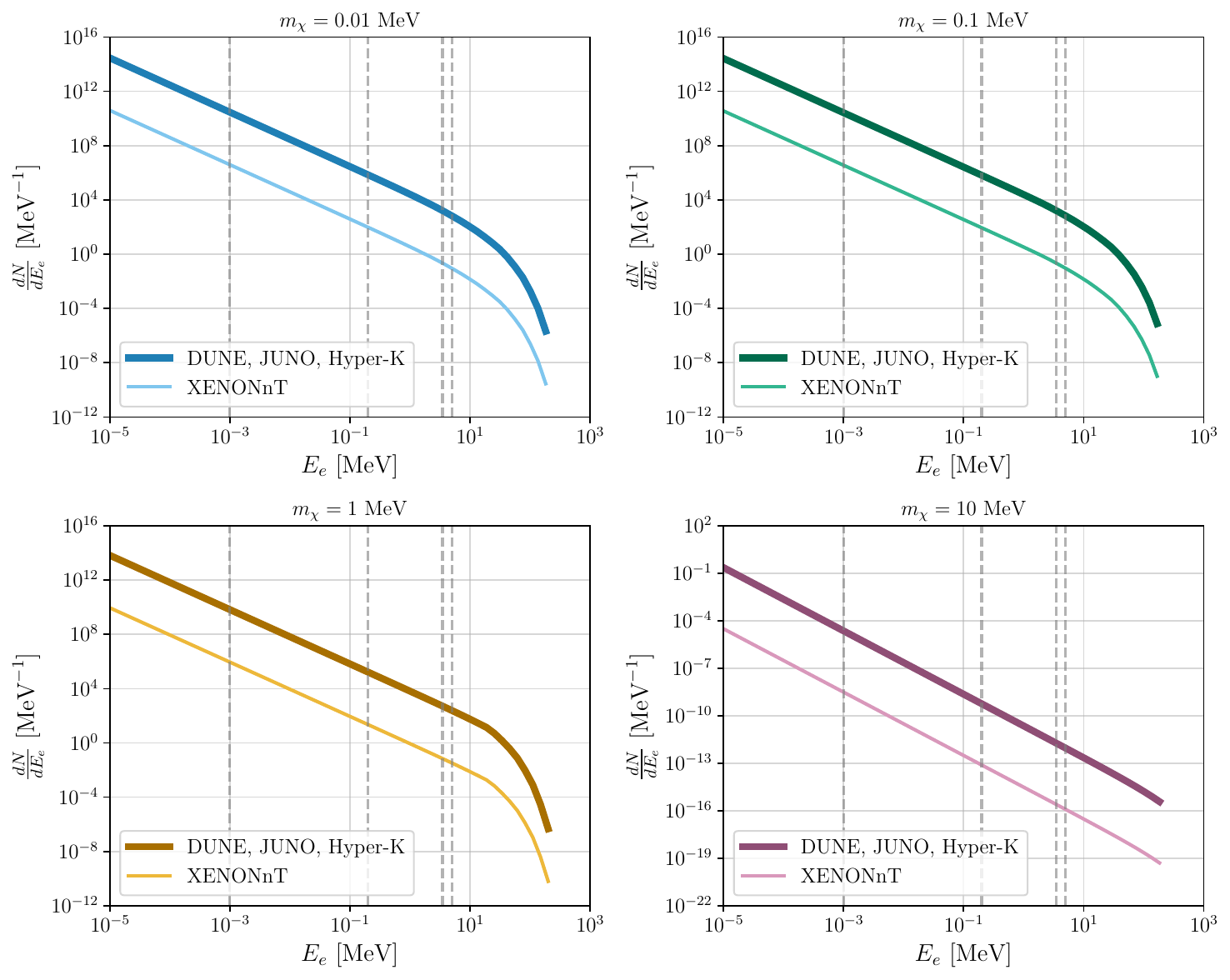}
        \caption{The time-integrated electron recoil spectra within the first year after the SN. Different experiments are considered for several values of $m_\chi$, $D = 1$~kpc and $\varepsilon = 10^{-9}$ are fixed. The four vertical dashed lines mark the lower electron-recoil energy thresholds of XENONnT, JUNO, Hyper-K, and DUNE, from left to right. The lower-right panel uses a different vertical-axis range because the plasmon-decay channel is absent for $m_\chi = 10$~MeV.}
    \label{fig:diff_event}
\end{figure*}

Figure~\ref{fig:diff_event} shows the time-integrated electron recoil spectra during the first year following the SN neutrino burst. The spectra are obtained by setting the lower integration limit in Eq.~\eqref{diff_rate} to the minimum effective mCP energy corresponding to an observation time of $\Delta t_{\mathrm{obs}}=1,\mathrm{year}$. Because the predicted spectra for DUNE, JUNO, and Hyper-K are nearly identical, they are shown together. Although XENONnT has a much smaller target mass, its substantially lower recoil-energy threshold allows it to probe lower values of $E_e$ that are inaccessible to the other three detectors. The vertical dashed lines indicate the electron-recoil energy thresholds of the four experiments.

\begin{figure*}[htbp]
    \centering
    \includegraphics[width=0.8\textwidth]{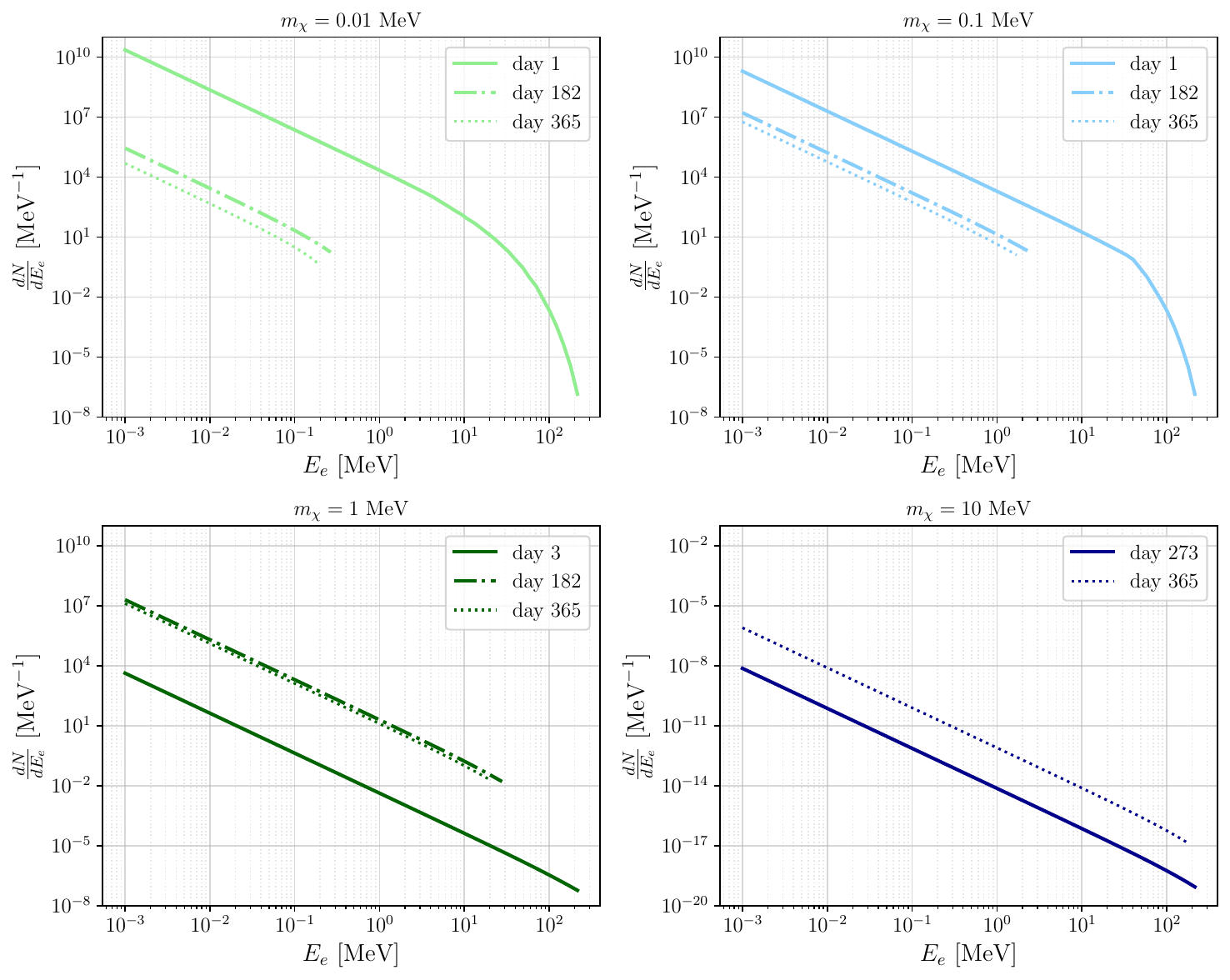}
    \caption{Electron recoil spectra at DUNE at different arrival times. Different panels represent different mCP masses. The solid curves correspond to the first one-day bin ($m_{\chi}$-dependent) associated with nonzero electron recoil. The dash-dotted curves show the spectra at midyear (day 182), and the dotted curves correspond to the last day of the year (day 365). Day $i$ denotes the time-delay interval $\Delta t \in [(i-1)~\mathrm{day},\, i~\mathrm{day}]$.}
    \label{fig:recoil_spectrum}
\end{figure*}

Figure~\ref{fig:recoil_spectrum} shows the electron-recoil spectra at DUNE in representative one-day arrival-time bins: the earliest bin containing mCP arrivals for each mass, a midyear bin, and the final day of the first year. For each one-day time bin, the upper integration limit in Eq.~\eqref{diff_rate} is set to the mCP energy whose arrival time coincides with the beginning of the bin.
The temporal evolution of the spectra distinguishes the mCP signal from both the prompt SN-neutrino burst and the approximately steady solar- and atmospheric-neutrino backgrounds, improving the prospects for mCP detection. Since this evolution depends on the mCP mass, the recoil spectra may also provide sensitivity to $m_{\chi}$. For $m_{\chi}=10~\mathrm{MeV}$, only two curves are shown because no mCP-induced electron recoils are expected during the midyear bin.

\subsection{Time Profile of the Detected Event Rate}
\label{app:time_profile}

\begin{figure}[htbp]
    \centering   \includegraphics[width=0.6\textwidth]{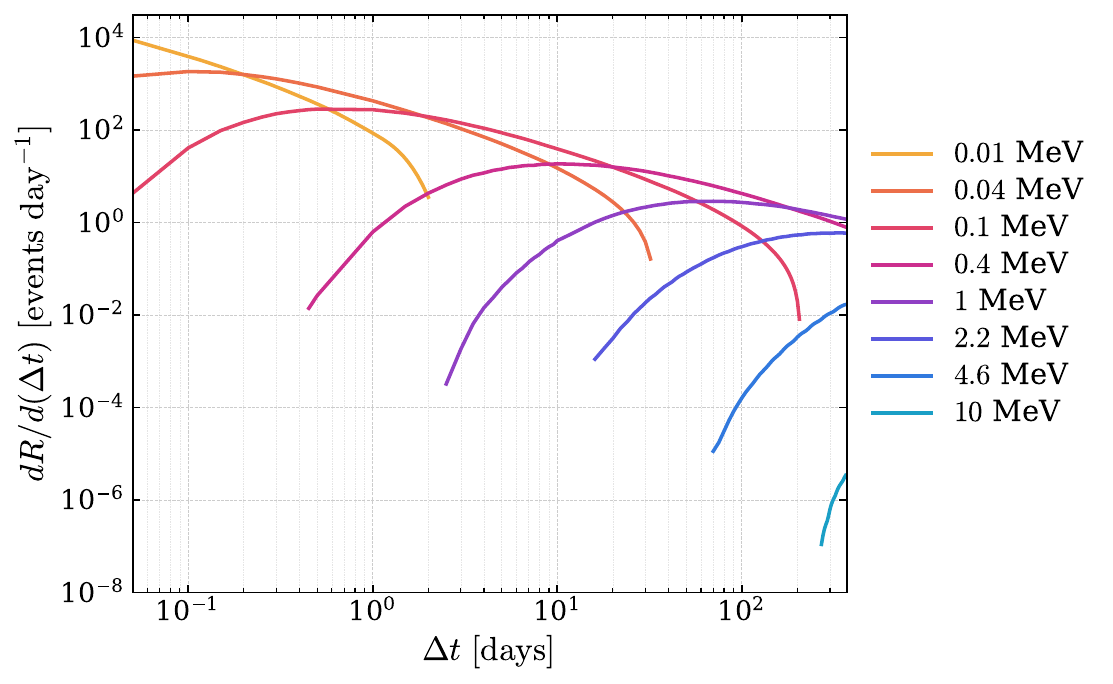}
    \caption{Time profile of the mCP-induced electron recoil event rate at DUNE as a function of ToF delay $\Delta t$ after a core-collapse supernova at $D = 1$~kpc, $\varepsilon = 10^{-9}$. Each curve corresponds to a fixed mCP mass $m_\chi$ ranging from 0.01 to 10~MeV, with both plasmon decay and $e^+e^-$ annihilation included. As $m_\chi$ increases, the peak shifts to later $\Delta t$ and the maximal rate decreases, as heavier mCPs travel more slowly at fixed energy.}
  \label{fig:time_profile}
\end{figure}

Fig.~\ref{fig:time_profile} shows the arrival-time profile, i.e., the differential event rate per target electron defined as:
\begin{equation}
\frac{1}{N_T}\frac{dR}{d(\Delta t)} =
\underbrace{\frac{d\phi_\chi}{dE_\chi}\bigg|_{\bar{E}_\chi(\Delta t)}}_{\text{flux at Earth}}
\times
\underbrace{\int_{E_e^{\rm thr}}^{E_e^{\max}(\bar{E}_\chi)} \frac{d\sigma}{dE_e}\,dE_e}_{\text{cross section}}
\times
\underbrace{\left|\frac{d\bar{E}_\chi}{d(\Delta t)}\right|}_{\text{Jacobian}}
\times
\underbrace{\Delta t_{\rm SN}}_{\text{burst duration}} .
\label{eq:time_profile_master}
\end{equation}
For a fixed $m_\chi$, the ToF relation in Eq.~\eqref{eq:time_delay} maps each arrival delay to a single incoming energy $\bar{E}_\chi(\Delta t)$. This value corresponds to the same timing cutoff that enters the effective lower limit in Eq.~\eqref{Emin_eff}:
\begin{equation}
\bar{E}_\chi(\Delta t) = m_\chi\,\frac{D + c\,\Delta t}{\sqrt{c\,\Delta t\,(2D + c\,\Delta t)}} ,
\label{eq:Echi_of_t}
\end{equation}
The flux at Earth, $d\phi_\chi/dE_\chi$ from Eq.~\eqref{diff_rate}, is evaluated at this energy. The supernova burst duration, $\Delta t_{\rm SN} = 10$~s, converts this flux into a fluence representing the number of mCPs per unit area per unit energy delivered during the neutrino burst. The Jacobian remaps the energy distribution into arrival time:
\begin{equation}
\left|\frac{d\bar{E}_\chi}{d(\Delta t)}\right| = \frac{m_\chi\, c\, D^2}{\left[c\,\Delta t\,(2D + c\,\Delta t)\right]^{3/2}} ,
\label{eq:jacobian}
\end{equation}
which follows from differentiating Eq.~\eqref{eq:time_delay}.

The cross-section factor integrates the elastic mCP-electron cross section $d\sigma/dE_e$ from Sec.~\ref{sec:detection} from the detector threshold $E_e^{\rm thr}$ up to the kinematic maximum recoil at a fixed incoming energy:
\begin{equation}
E_e^{\max}(E_\chi) = \frac{2 m_e \left(E_\chi^2 - m_\chi^2\right)}{m_e^2 + m_\chi^2 + 2 m_e E_\chi} .
\label{eq:Eemax}
\end{equation}
For a fixed incoming mCP energy $E_\chi$, Eq.~\eqref{eq:Eemax} determines the maximum allowed electron recoil energy. Conversely, solving $E_e^{\max}(E_\chi) = E_e^{\rm thr}$ yields the minimum incoming mCP energy $E_{\rm min}$ required to produce a recoil above threshold, as defined in Sec.~\ref{sec:detection}.

The four factors carry units of $[\mathrm{cm^{-2}\,s^{-1}\,MeV^{-1}}] \times [\mathrm{cm^2}] \times [\mathrm{MeV\,s^{-1}}] \times [\mathrm{s}] = [\mathrm{s^{-1}}]$ per target electron. Multiplying by $N_T$ and converting via $86400~\mathrm{s/day}$ yields the detected event rate per day, $dR/d(\Delta t)$, which Fig.~\ref{fig:time_profile} displays. The $1/N_T$ normalization removes the detector-specific target size. Equation~\eqref{eq:time_profile_master} serves as the time-domain counterpart to Eq.~\eqref{diff_rate}. It utilizes the identical flux and cross section, but fixes $E_\chi$ by $\Delta t$ and applies the Jacobian map to arrival time instead of integrating over $E_\chi$ at a fixed $E_e$.

\clearpage
\newpage

\section{mCP Signals from Past Supernovae}
\label{app:past_SNe}

Historical observations of nearby supernovae, such as SN1987A ($D \simeq 50$~kpc) in the Large Magellanic Cloud and the Galactic SN1054 ($D \simeq 2$~kpc), raise two natural questions: whether mCP events were recorded at the time of the explosion, and whether they remain detectable today. This appendix addresses both questions. The analysis demonstrates that these past supernovae leave no observable signature in the current search channel.

\subsection{Detection at the Time of Explosion}
\label{app:past_SNe_then}

Neglecting the additional $D$-dependence from the finite ToF observation window, the total event count scales relative to the reference values $D_0 = 1$~kpc and $\varepsilon_0 = 10^{-9}$ as:
\begin{equation}
N(D,\varepsilon) \simeq
N(D_0,\varepsilon_0)
\left(\frac{\varepsilon}{\varepsilon_0}\right)^4
\left(\frac{D_0}{D}\right)^2 .
\label{eq:N_scaling}
\end{equation}
The two factors of $\varepsilon^2$ originate from mCP production in the SN core and mCP-electron scattering in the detector, whereas the $D^{-2}$ factor accounts for geometric dilution. For a fixed number of events and fixed $m_\chi$, Eq.~\eqref{eq:N_scaling} yields
\begin{equation}
\varepsilon_{\min}(D) 
= \varepsilon_{\min}(D_{0}) \left(\frac{D}{D_{0}}\right)^{1/2},
\label{eq:eps_scaling}
\end{equation} 
provided the ToF-induced shift in the energy integral remains small.

The maximum supernova distance $D_{\max}$ at which a detector probes below the SN1987A cooling bound derives from the intersection of the full numerical sensitivity curve with the cooling-bound curve, rather than a direct extrapolation of Eq.~\eqref{eq:eps_scaling}. Among the facilities evaluated, the large-scale neutrino detectors provide the maximum spatial reach. The numerical pipeline from Sec.~\ref{sec:results} yields $D_{\max} \simeq 18$, $21$, $41$, and $83$~kpc for DUNE, Hyper-K, XENONnT, and JUNO, respectively. DUNE and Hyper-K extend the discovery reach below the cooling bound across most of the Galactic disk, whereas XENONnT and JUNO extend beyond the Galactic disk, with JUNO reaching Magellanic-Cloud scales.

A JUNO-scale detector operating during SN1987A would have recorded tens of mCP-induced electron-recoil events in the first year after the burst for $\varepsilon = 10^{-9}$. Because mCPs remain ultra-relativistic at core temperatures of $30$~MeV, the high-energy flux arrives within years to decades of the core collapse. However, SN1987A predates modern detection capabilities. For example, Kamiokande-II possessed roughly $1\%$ of the JUNO target mass, lacked the necessary low-threshold electron-recoil sensitivity, and operated without a timing-triggered delayed-event pipeline. Furthermore, no dedicated retroactive searches for mCP-induced events from SN1987A exist in current detector data.

\subsection{Detection Today}
\label{app:past_SNe_now}

The second question concerns whether mCPs from past supernovae continue to arrive in detectable quantities today. The ToF delay relation in Eq.~\ref{eq:time_delay} maps a one-year observation window to a narrow mCP energy slice for fixed $m_\chi$ and $D$. For long elapsed times $\Delta t \gg 1$~yr, the relative width of this slice scales approximately as $1/(2\Delta t)$. This scaling yields a width of $\sim 1\%$ for SN1987A ($\Delta t \simeq 39$~yr) and $\sim 0.05\%$ for SN1054 ($\Delta t \simeq 970$~yr). Because the mCP spectrum falls steeply with energy, these narrow slices contain insufficient particle fluxes to produce detectable event rates today.

\end{document}